# Patterns of Text Reuse in a Scientific Corpus

**Daniel T. Citron** [*], **Paul Ginsparg** [*†]

[*] Dept of Physics, Cornell University, and [†] Dept of Information Science, Cornell University

**We consider the incidence of test "reuse" by researchers, via a systematic pairwise comparison of the text content of all articles deposited to arXiv.org from 1991–2012. We measure the global frequencies of three classes of text reuse, and measure how chronic text reuse is distributed among authors in the dataset. We infer a baseline for accepted practice, perhaps surprisingly permissive compared with other societal contexts, and a clearly delineated set of aberrant authors. We find a negative correlation between the amount of reused text in an article and its influence, as measured by subsequent citations. Finally, we consider the distribution of countries of origin of articles containing large amounts of reused text.**

## 1. Introduction

Detection of text reuse in large corpora has been facilitated in recent years by increasingly sophisticated algorithms, more powerful hardware, and more widespread availability of texts. In [1], the winnowing methodology of [2] was adapted to consider text reuse in a scholarly corpus. In the present work, we use refined heuristics to perform a more systematic assessment on a much larger corpus, and to look for patterns in text reuse. As our dataset, we use the texts contained in the arXiv, a repository of articles deposited by researchers in Physics, Mathematics, Computer Science, and some related fields available at http://arxiv.org/ .[1] The dataset used in the current analysis consists of roughly 757,000 articles from mid-1991 to mid-2012, towards the end of which time the repository was receiving roughly 80,000 new submissions per year.[2]

One motivation for undertaking this analysis of arXiv data was the known incidence of text copying and plagiarism, usually noticed by readers, and sometimes reported in the news media. The authors of [4], for example, pointed out unattributed use of their text in a series of four arXiv articles in 1999. A news article from 2003 [5] described the case of an unknown author who tried to establish research credentials by submitting texts largely copied from other sources. In a 2007 news article [6] discussing the earlier version [1] of the work here, it was noted that the cases detected spanned a wide range, from 27 pages of lecture notes by another author used verbatim in a thesis, to reuse of common introductory material, to text overlaps of benign common phrases. Shortly afterwards, as reported in another news article [7], a large number of articles from a group of coauthors was withdrawn due to reuse of text copied from a variety of sources.

Practical considerations for running the arXiv site provide another motivation, since problematic authors can inconvenience readers by producing more than their share of articles, reusing of large blocks of their own text. Screening for this had been haphazard, and moreover a systematic baseline to identify outliers, and to provide a principled response to the claim "this is common practice, everyone does it," was needed. The current work, re-employing the methodology of [1], gives a more systematic assessment of the statistics of text reuse in the arXiv dataset, and permits identification of the extremes of the distribution, so that outliers can be publicly flagged.[3]

While there is no universal standard pertaining to reuse of text in scientific publications, many universities and publishers have established explicit guidelines and provide training (e.g., [8, 9, 10]). In Appendix B, we provide a brief survey of representative policies.[4] Journal publishers provide effective international guidelines, and the American Physical Society's, for example, are unequivocal regarding text reuse [11]: "Authors may not ... incorporate without attribution text from another work (by themselves or others), even when summarizing past results or background material." We will see that arXiv submissions do not always conform to these exacting standards, and yet are published by journals, indicating that editors do not systematically employ an automated screen.

To be clear, we are careful in what follows to restrict attention to simple text overlaps. We make no attempt to detect "plagiarism" in its most general form, which includes unattributed use of ideas (whether or not text is copied).[5] We also make no attempt to detect text copied from sources outside of arXiv (legacy print material, Wikipedia, the rest of the WorldWideWeb, etc.), so our focus is further restricted to a simple factual statement regarding textual overlap of materials within arXiv.[6] Since our intent is to establish a baseline for existing community behavior, the presentation in this article identifies no authors.

## 2. Methodology

We have preprocessed some features specific to the arXiv texts to help eliminate false positives. The reference sections are removed from the texts, since overlaps among references can be ignored. We have also tried to identify blocks of text in quo-

**Significance**

**In the modern electronic format, it is both easier to reuse text and easier to detect reused text. This is the first comprehensive study of patterns of text reuse within the full texts of an important large scientific corpus, covering a twenty year timeframe. It provides an important baseline for what is regarded as standard practice within the affected research communities, a standard somewhat more lenient than currently applied to journalists, popular authors, and public figures.**

**Reserved for Publication Footnotes**

[1] Now administered by the Cornell University library. For some recent informal histories, see [3].

[2] See http://arxiv.org/help/stats/2012_by_area/index

[3] This policy was implemented in the summer of 2011, see http://arxiv.org/help/overlap

[4] All appendices are included in the Supplementary Materials.

[5] In [12], it is argued that the most severe offense is unattributed use of ideas from non-publicly available documents, such as grant proposals.

[6] Commercial resources, such as Ithenticate, use a much larger dataset. See in particular CrossCheck [13], implementing Ithenticate for research publications, and used by member publishers to screen journal submissions [14]. That coverage is still far from as comprehensive as available via commercial search engines, as assessed by comparing to results from the Google custom search API.

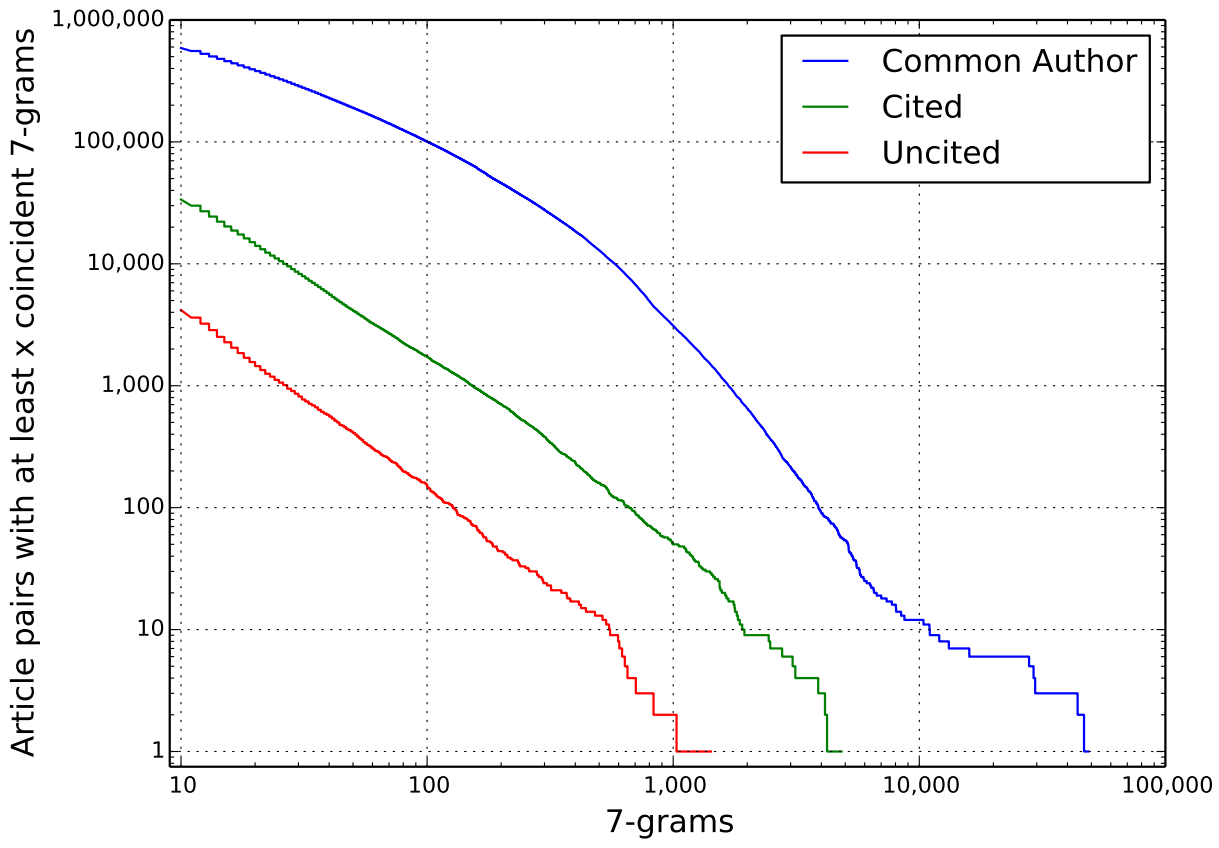

**Fig. 1.** Cumulative distribution of overlapping 7-grams for article pairs with Common Author in blue (upper curve), Cited in green (middle), and Uncited in red (lower). The vertical axis is the number of article pairs with at least the number of overlapping 7-grams given on the horizontal axis (starting with a minimum of at least 10). Both horizontal and vertical axes are logarithmic.

tation marks where possible (but find in any event that block quotes comprise a tiny fraction of the overlaps in the corpus). For the purposes of this analysis, we have also excluded articles from very large experimental collaborations, since the long lists of author names (and other boilerplate) can masquerade as authors reusing their own text.

To detect text overlaps between arbitrary pairs of articles efficiently, we employ an extension of the methodology described in [1], as adapted from [2].[7] Each article can be effectively "fingerprinted", with its content represented by a set of hashes stored in a database that resides in RAM for rapid lookups. The hashes are determined by sequences of seven words in the article, called 7-grams, eliminating sensitivity to commonly used shorter sequences (e.g., "this article is organized as follows"). The number of hashes retained for each document are "winnowed" [2] (reducing their number by a factor of 3.6 at a small loss of sensitivity to words sequences of less than 12 words), and further reduced (by another 4%) by eliminating "common" 7-grams [1]. The resulting hash database requires about 12Gb of RAM, and permits many hundreds of lookups per second on inexpensive hardware.

In the remainder of this article, "7-grams" will refer to the winnowed uncommon 7-grams harvested using the winnowing methodology described in Appendix A. For typical amounts of text overlap, the number of overlapping words is roughly six or seven times the number of such overlapping 7-grams. Thus two articles with 100 overlapping 7-grams can be thought of as having roughly 35 sentences in common.

## 3. Aggregate measures of text reuse

In measuring rates of text overlap, we distinguish three modes of reuse, in increasing order of severity: we use "Common Author" (AU) to designate a pair of overlapping articles with at least one author in common; "Cited" (CI) to designate a pair with no common authors but at least one article cites the other; and "Uncited" (UN) to designate a pair with neither common authors nor citation of the earlier article. Fig. 1 shows the results of this analysis for the roughly 757,000 articles in the database in the summer of 2012 (accumulated since 1991), consistent with, and updating and refining the results of [1]. Each of the three curves represents the cumulative number of article pairs with at least the number of coincident 7-grams specified on the horizontal axis, with AU, CI, and UN modes depicted in blue, green, and red, respectively. For example, the AU curve (blue) indicates roughly 100,000 cases with at least 100 7-grams in common, 3000 with at least 1000 in common, and only about 10 such pairs with as many as 10,000 in common.[8] The CI curve (green) ranges from the tens of thousands of pairs for ten 7-grams in common, down to a handful of pairs having a few thousand 7-grams in common. The UN line (red), for article pairs with neither authors in common nor citation, ranges from thousands of article pairs with ten 7-grams in common to ten pairs with at least 500 in common. We see from the log scale in the figure that AU text reuse is approximately an order of magnitude more frequent than CI text reuse, and approximately two orders of magnitude more frequent than UN text reuse.

At first glance, the data represented in fig. 1 suggests significant cause for concern: is the literature[9] really so replete with text reuse? Do so many authors really repurpose their own text and that of other authors, with or without attribution? Before jumping to conclusions, we should consider various mitigating circumstances.

In the case of authors reusing their own past material, it may be that such recycling is sometimes "acceptable" practice. For example, doctoral theses in physics once consisted largely of original materials, but graduate students are now expected to publish multiple articles, and it is a common practice for the thesis to incorporate some of these articles in their entirety, without changes. Similarly, in most disciplines it is considered acceptable to have separate short and in-depth versions of the same work, with the former incorporated into the latter.

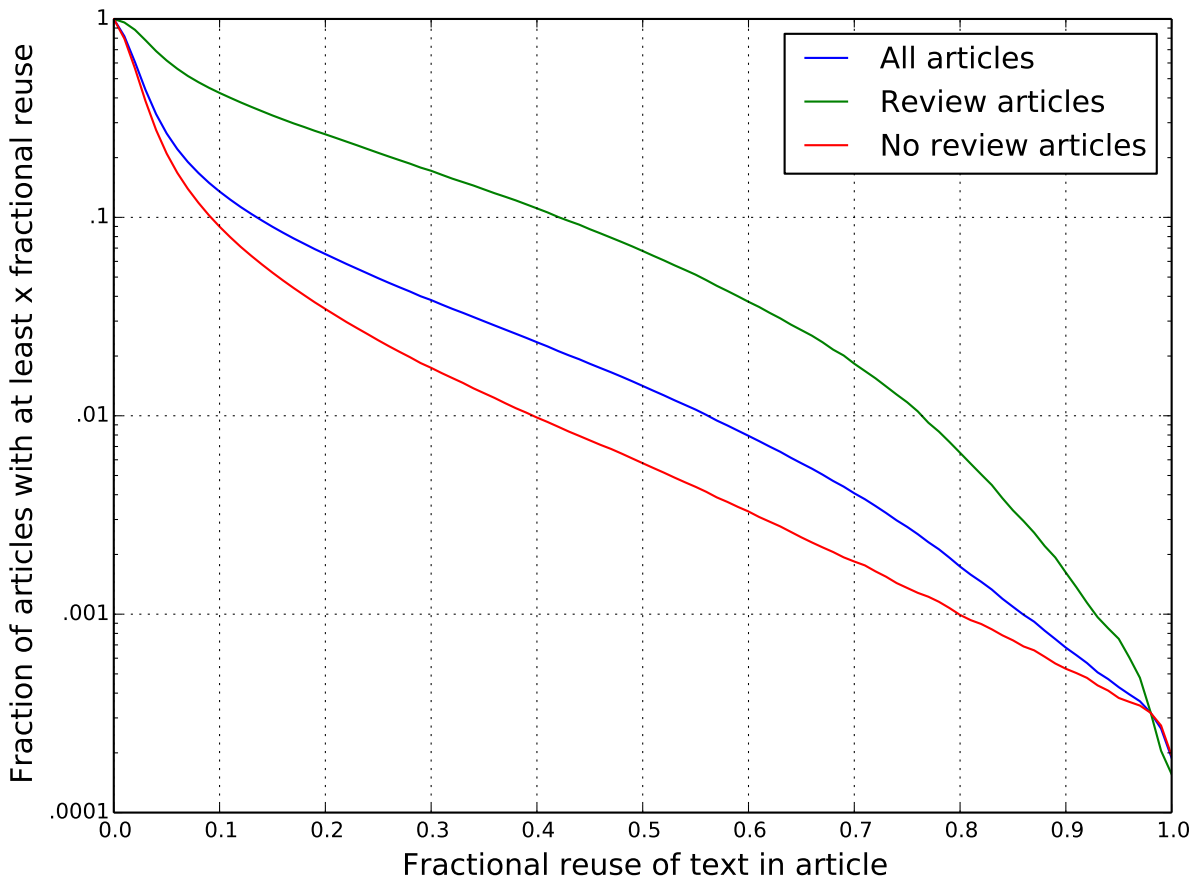

**Fig. 2.** The vertical axis gives the fraction of articles with at least the indicated fraction of reused 7-grams on the horizontal, where green (upper) signifies Review, red (lower) signifies non-Review and blue (middle) combines both. The vertical is plotted on a log scale to permit seeing the full range; the dropoff in fraction of articles with given amount of reuse would be much steeper on a linear scale.

[7] We summarize the procedure here, and provide more technical details in Appendix A.

[8] The number of article pairs with at least 10 or more 7-grams in common is of order 600k, about 2 per million of the total possible $(757k)^2/2 \approx 278B$ total article pairs.

[9] Recall that the vast majority of arXiv submissions appear in the conventional peer-reviewed literature, with the primary exceptions being theses, conference proceedings, lectures, and other "review-type" materials discussed earlier (and excluded from subsequent analysis).

[10] Review articles pose an additional challenge, since standard software used to include pdf figures from other articles sometimes carries along "hidden" text surrounding the figure from its original context, invisible to the author and reader in the new context, but nonetheless seen by the pdf to text converter and flagged as a large text overlap.

[11] Nonetheless a study of seven million biomedical abstracts [15] suggests that redundant publication has been increasing in those areas.

Another perhaps more contentious case is that of review articles. Some authors take it for granted that review articles should be original syntheses of past work, whereas others feel free to use large blocks of material from earlier articles.[10] Lecture notes, book contributions, and other popularizations constitute another form of publication in which liberal reuse of earlier material could be considered acceptable.

Attitudes towards reuse of text in conference proceedings also vary widely, differing between authors and fields. In Physics, for example, conferences are a secondary publication venue with little prestige, and it is accepted that material is recycled from earlier articles. In Computer Science, on the other hand, conference publication is a primary venue, and significant self-copying by authors is not the norm. In Mathematics, it is considered standard practice to restate important theorems or definitions from one's own work (or from elsewhere, with attribution). In many disciplines, it is standard practice to reuse blocks of material describing experimental facilities or procedures.

To assess the extent to which text reuse is concentrated among articles in the above classes (review articles, conference proceedings, dissertations, etc.), we harvested from the article metadata keywords such as "review", "proceedings", "thesis", etc., to detect submissions that were self-identified by submitters as review-type. We designate these articles as "Review", and partition the results from fig. 1, accordingly, in fig. 2. The horizontal axis in the figure shows the *fractional* text reuse within the article, given by the fraction of 7-grams in an article that appear in some other article, and the vertical axis indicates the fraction of articles in the database with that percentage of reuse. The middle solid line (blue) in the figure shows the fraction of *all* articles with at least the indicated fractional reuse, so for example roughly 2% of the database consists of articles 50% of whose 7-grams appear elsewhere. The upper solid line (green) isolates from that set the fraction of articles self-identified in the Review category, and provides the fraction of those articles with the indicated fractional reuse. We see that roughly 7% of those articles contain at least 50% reuse, whereas less than .6% of the non-Review articles (solid red line) have that much text reuse. Fig. 2 indicates that the vast majority of the AU text reuse in fig. 1 occurs in contexts generally regarded as acceptable by the community.[11] The solid red curve depicts a non-negligible percentage of text reuse that occurs outside of those contexts.

Given the prevalence of text reuse, it is natural to wonder how these texts are distributed among authors: Is it concentrated among a few serial offenders or distributed more widely? Are the authors prominent or obscure? Do the texts in question have significant impact? In the following sections, we apply further filters to the data to address these questions.

## 4. Author-specific measures

We first turn to the question of how the aberrant text reuse is distributed among authors: is it all of the authors some of the time, or some of the authors all of the time? It might, for example, be reassuring if only a relatively small group of authors is responsible for the majority of observed cases. Here we characterize the extent to which text reuse is "normal" behavior, to be able identify the abnormal behaviors.

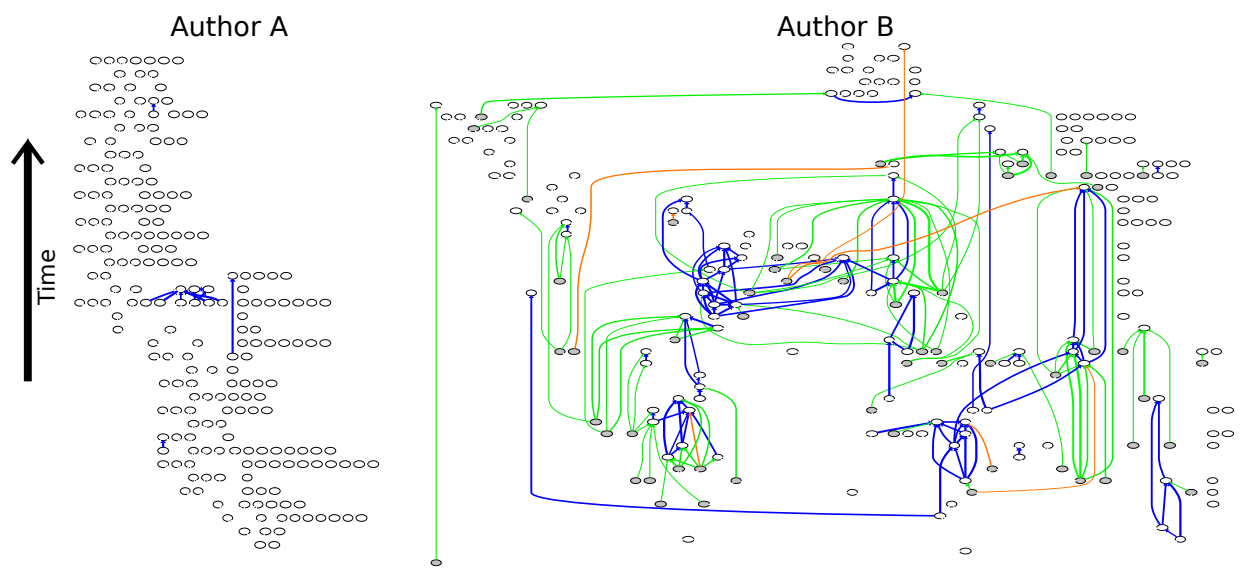

**Fig. 3.** Examples of text overlap networks of two authors A, B. The colors are as in fig. 1, with blue, green, and red edges representing AU, CI, and UN overlaps, respectively. Edge thickness is proportional to article overlap. Articles are arranged by time of submission, with earlier at bottom and more recent at the top. Uncolored nodes are texts coauthored by the author in question, and gray nodes are texts by other authors.

**Text overlap networks.** In general, a network is a collection of nodes linked together by edges, where each node represents an object and each edge between two nodes represents a connection or relationship between the corresponding objects. Here we introduce a text overlap network, in which each node represents an article and each edge a pairwise textual overlap between articles. Because articles published later in time copy from earlier ones (and not vice versa), all edges in the network are directed forward in time to represent text transfer. Each edge is weighted according to the number of 7-grams that the two connected articles have in common, and edges are color-coded blue, green, and red, resp., to indicate AU, CI, and UN modes of text overlap.

In the text overlap network for articles written by a specific author, the density of edges is proportional to the amount of reused text, so the network provides a useful visualization of text reuse, and for assessing overlaps among articles by a particular author or group of authors. Fig. 3 shows the text overlap networks of two authors with vastly different patterns of text reuse. Articles by Author A have few overlaps: of 217 co-authored articles, only 6 contain previously published text; whereas Author B's text overlap network is far more densely connected. The blue edges reveal clusters of articles by that author with material copied from one another. Furthermore, in contrast to Author A, Author B has also reused text from articles written by other authors (represented by green- and red- colored edges.)

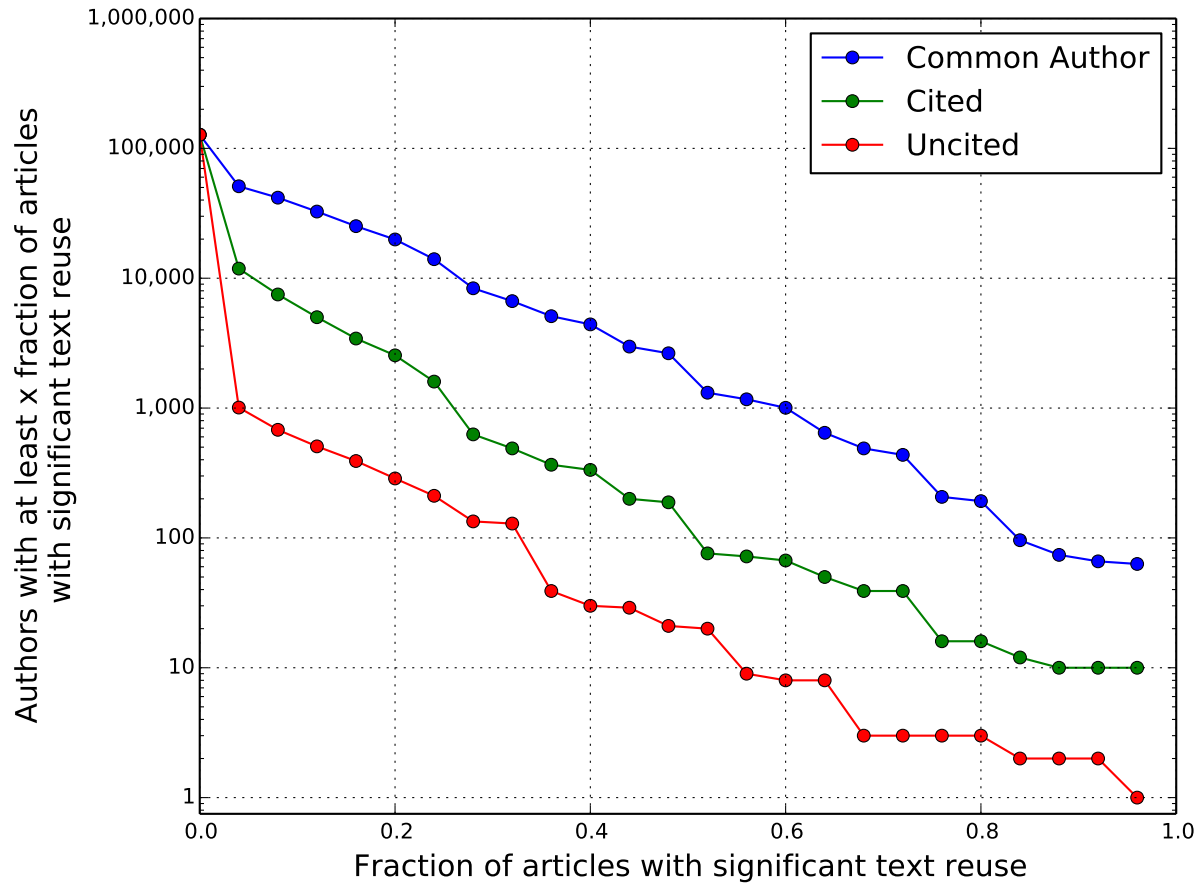

**Fig. 4.** Cumulative histogram of the number of authors (vertical axis) having at least a given fraction of their articles with significant text overlaps (horizontal axis). The data is restricted to authors with at least four articles in the corpus. For example, roughly 1000 such authors have significant AU text overlap (upper line) in at least 60% of their articles. (Review-type articles, as described in the lead-up to fig. 2, are excluded from this data.)

While it is possible to produce large numbers of articles more quickly by copying from prior content, Author A in fig. 3 illustrates that a large number can also be generated without such copying. Author A submitted 177 articles and Author B submitted 174 articles between January 2000 and June 2012, each averaging about 1.2 articles submitted per month in that period, but only the latter author habitually copied previous text. Not all prolific authors are habitual text reusers, nor are all text reusers necessarily as prolific as Author B. But while many or most authors have little desire to retread the same material more than once, preferring to move on to fresh material, there are authors whose publications tend to consist largely of previously published material, with minimal new content. In sec. 5, we will consider the extent to which such text reuse is correlated with subsequent citations.

Appendix C provides a few more samples of overlap networks for authors with very high frequencies of text reuse, and appendix D provides examples of text overlaps that can be difficult to classify.

**Detecting serial copiers.** To quantify an author's tendency to reuse text, we consider the fraction of an author's articles that are derivative, i.e., include more than a specified threshold of copied material. To focus on the more significant instances of text overlap, we consider only cases of at least 100 7-grams in the case of AU overlaps, and at least 20 shared 7-grams in the case of CI or UN overlaps. Recalling the winnowing procedure, these thresholds correspond approximately to 35 and 7 sentences of copied text, respectively. The lower thresholds for CI and UN overlaps reflect their lower frequencies relative to AU overlaps. Our results are insensitive to the choice of thresholds in the sense that the same behavior from the same groups of authors is flagged for a range around these values. The thresholds also reduce false positives resulting from artifacts of pdf to text conversion, mis-characterized author or citation lists, restatement of theorems, or an occasional block quotation of text.

To restrict attention to *habitual* reuse of text, we include only authors who appear on at least 4 articles. Fig. 4 shows a cumulative histogram of the number of authors whose articles contain a given fraction of significant AU, CI, and UN text overlaps. For example, an author with ten articles, four of which have significant AU overlap, would contribute to the upper (blue) line for x-axis values less than or equal to .4. Most importantly, we see that the number of authors with articles flagged for each of the three types of overlaps drops significantly as the fraction of problematic articles increases from 0%. Of the total 127,270 authors in the dataset, only 51,060, 11,860, and 1010 have more than 4% of their articles contain AU, CI, and UN text overlaps, resp. The vast majority of authors, therefore, either never or only rarely reuse significant amounts of text in new publications. In the more problematic region, we see only 14,020, 1600, and 210 with at least 24% of their articles containing significant AU, CI, and UN overlaps, resp. We infer that the practice of reusing text is uncommon and is restricted to a minority of serial offenders, responsible for the heavy tail in fig. 1.

## 5. More author sociology

**Text overlap and citations.** Having seen that the problematic behavior is restricted to a small minority of authors, we turn to assess the impact of their work. We use the number of citations that each article has received as a proxy for its influence, and investigate any correlation with the amount of copied content in the article. We focus on a subset of 116,490 articles for which we have relatively clean citation data, primarily in Astrophysics and High Energy Physics.[12] The articles selected for this subset appeared prior to the start of 2011, giving them time to accumulate citations. To provide a better proxy for an article's real influence, we discard self-citations, i.e., citations by articles with any coincident authors.

We estimate the fraction of copied content in an article by dividing the number of 7-grams that have appeared previously by the total number of 7-grams from the article, without removing the common 7-grams. Retaining both common and uncommon 7-grams in this instance gives a better measure of the extent to which authors rely on earlier texts. We exclude from the dataset all articles with 95% or more overlap with other articles, since these are typically articles erroneously submitted more than once to arXiv after minor revisions, and are not the type of overlap at issue.[13] We also exclude from the dataset all articles containing less than 5% reused content, since these signify likely failure of the pdf to text conversion, for example due to font issues, making the estimate of fraction of copied content unreliable.[14] We have also excluded Review-type articles (conference proceedings, theses, etc., as described in the lead-up to fig. 2) to avoid creating an artifactual correlation between reused content and low number of citations.

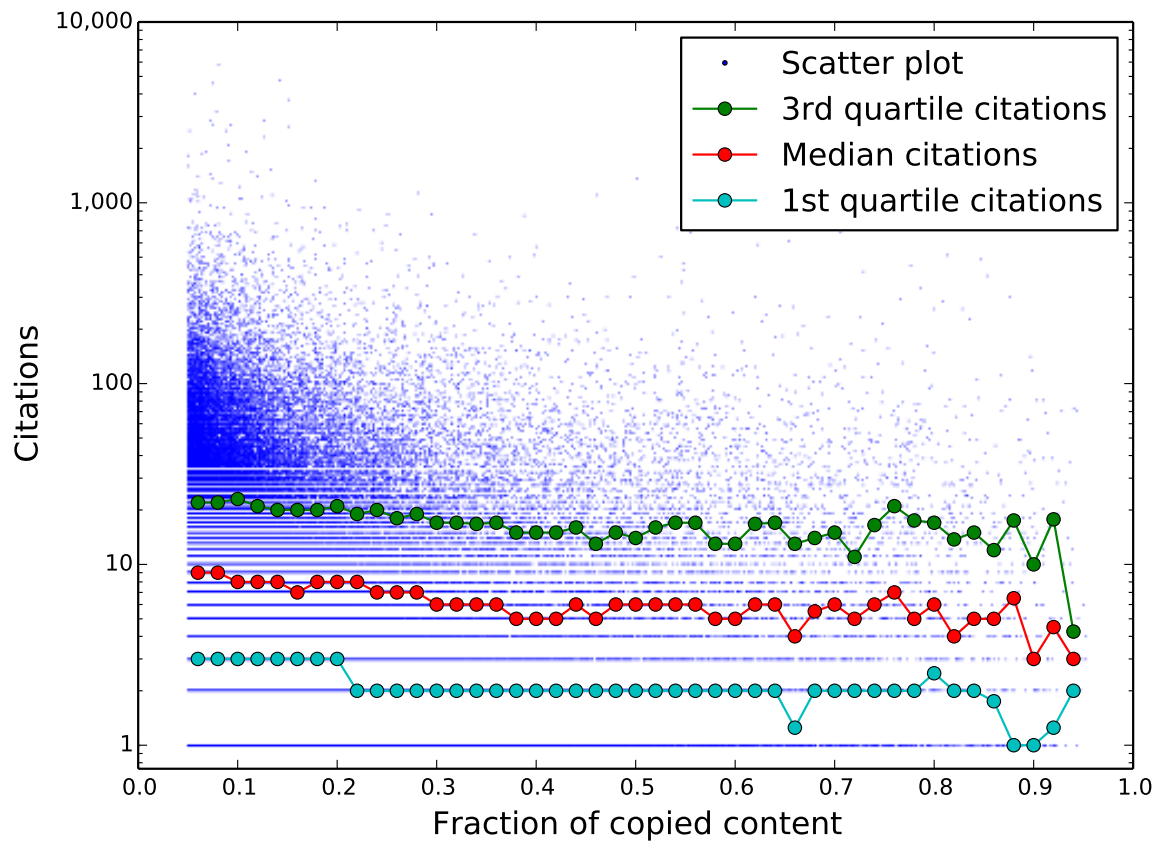

**Fig. 5.** Scatter plot of the number of citations vs. fraction of copied content (blue). The median number of citations vs. fraction of copied content is shown in red (middle line of points), indicating a negative correlation, with Spearman correlation coefficient $r = -.739$ ($p = 6.76 \cdot 10^{-9}$). The y-axis is logarithmic, and the plot also shows 1st and third quartiles for the citations.

[12] Thanks to Alberto Accomazzi for providing citation data from the Astrophysics Data System.

[13] This happened historically when users inadvertently created a submission with a new identifier rather than using the replace function to create a new version of an existing submission, with the same identifier. This problem has been largely eliminated by the daily overlap screening, with submitters now instructed to replace an existing submission if excessive overlap is detected.

[14] Since we are retaining as well the "common" 7-grams for this purpose, all properly converted texts will now exhibit some reused content.

Fig. 5 shows the number of citations plotted against the fraction of copied content contained in each article. The wedge of points at the left of the scatter plot shows that there is a higher variance in the number of citations for articles containing low amounts of copied content. Qualitatively speaking, it is more likely for articles with a low fraction of copied content to receive very many citations, whereas it is relatively rare for articles with a high fraction of copied content to receive the same number of citations. To quantify this, we also plot the median number of citations as a function of fraction of copied content in red, and calculate a Spearman correlation coefficient of $r = -.739$ ($p = 6.76 \cdot 10^{-9}$). This illustrates a strong decreasing trend of citations for articles with increasing copied content. The presence or absence of reused text in an article thus serves as an artifactual quality flag, with articles having large amounts of unoriginal content cited less frequently.

Since the articles are less frequently cited and presumably little read, it is tempting to speculate that the reused content in these articles goes largely unnoticed and undetected. Another reason that text reuse might go undetected is that the articles from which the text is copied are also less well-read. In Appendix E, we present data showing that there is as well a negative correlation between the amount of reused content *from* an article and the number of citations that article received, even after screening for author self-copying. This may result from authors working in overall less active subject areas (e.g., [7]); or may be due to a tendency for authors to borrow text from authors of the same nationality even in more active research areas, where articles by authors of that nationality are already correlated with fewer citations.

**Author demography.** We now investigate whether articles containing large amounts of reused text come from a uniform distribution of countries, or only from some more restricted set. In order to submit, authors must register an email address with arXiv, and we assign a country of origin to each article using the country code associated with the email address of the submitting author. (Note that we shall ignore any subtleties associated with multinational collaborations.)

We have employed two methods here to measure the amount of copied content in an article: either by estimating the total fraction of reused content, or by using a link measure based on the number of articles that have at least 100 7-grams in common with articles by the same authors (or at least 20 in common with articles by different authors). There are many more ways to rate countries in aggregate, e.g., using the percentage of copied articles by either of the above metrics, or the percentage of authors with more than some threshold of flagged articles, etc. Our intent here is not to find some quantitative means of rating different countries' research output or to flag individuals or nationalities for unethical behavior — it is only to give a flavor for the demographics of the authors involved. Shedding light on the nature of the problem may help to address it.

Labeling the articles by country of origin according to email domain, we first use the fraction of copied content in each article, as in fig. 4. We ignore countries from which fewer than 40 articles have been submitted, as providing insufficient data to resolve a clear pattern. Setting the threshold for flagging articles at 20% reused text identifies a group of countries with more than 15% of their submissions flagged. For comparison, under 5% of submissions from the United States and United Kingdom and under 10% of submissions from China, Turkey, and India were flagged by this criterion. Increasing the flagging threshold to articles with at least 50% reused content gives roughly the same group of countries with over 5% of their submissions flagged. For comparison, fewer than 1% of submissions from the US and UK are flagged by this criterion. Using an alternate criterion of more than 100 7-grams for AU, and more than 20 for CI or UN text reuse, again roughly the same group of countries appears with more than 11% of their articles flagged for text overlaps. With this criterion, less than 4% of articles from the US and UK are flagged, and China, Turkey, and India have 7-8% of their submissions flagged.

The countries that consistently, regardless of metric, contain the highest percentages of flagged submissions are (listed alphabetically): Bangladesh, Belarus, Bulgaria, Colombia, Cyprus, Egypt, Iran, Jordan, Kazakhstan, Kyrgyzstan, Latvia, Luxembourg, Micronesia, Moldova, Pakistan, Saudi Arabia, Uzbekistan. To screen for countries dominated by a small group of highly prolific authors, we also consider countries with high percentages of problematic submissions that had at least 100 distinct submitting authors. This group of countries includes (again, listed alphabetically): Armenia, Bulgaria, Belarus, Colombia, Egypt, Georgia, Greece, Iran, Romania. The exact order of these lists depends on which of the metrics is used, but we emphasize that the specific ordering is unimportant. The dataset is small in some cases, and the sample may be strongly biased by the subset of researchers most likely to upload to arXiv. Even so, the clear signal according to these criteria is that articles from developing countries where English is not widely spoken tend to contain large amounts of reused text at a much higher rate than the norm.[15] The practices may have developed due to differences in academic infrastructure and mentoring, or incentives that emphasize quantity of publication over quality. The Internet provides unprecedented global access to research-related materials and guidelines, so targeted supplementary resources might help ameliorate educational and cultural gaps.

A concern raised by the discussion above is that the negative correlation between citations and copied text seen in fig. 5 may be biased by country of origin, if researchers from certain countries tend to produce articles with large amounts of copied text, and articles from that country tend to receive few citations. In Appendix F, we screen for this effect and find that the negative correlation persists for countries with relatively low rates of copied text.

## 6. Observations

**Experience.** Starting in June 2011, submissions to arXiv have been marked with an "admin note", indicating text overlap with other arXiv submissions. The note is added to the "Comments" line in the submission's metadata, and is visible to all readers when the submission is announced. Roughly 250 submissions per month are currently flagged, corresponding to just over 3% of new submissions daily. They are flagged according to the methodology described in this article, as AU, CI, or UN text overlap, when the amount is well above the statistical background level for the respective types. The added notes are simple factual statements regarding relatively unambiguous textual overlap of materials within arXiv.[16] They are informational to readers, who may find it useful to know when an article draws heavily from another. They can also be informative to authors from different educational backgrounds, unaware that importing large sections of text from their earlier articles, or from articles by others, is not common practice.

[15] This is consistent with the results of [16], which found an association between retractions for plagiarism in the medical literature with first authors affiliated to lower-income countries.

[16] As discussed earlier, there is no systematic scan for text copied from sources outside of arXiv, and no attempt to detect "plagiarism" as more generally defined, as unattributed use of ideas independent of copied text. The exceptions described earlier for review articles, theses, conference proceedings, book contributions, multi-part articles, and so on, are respected, so that common-authored overlaps are not flagged in cases that appear to be accepted as common practice.

The reaction of authors has fallen into three classes: a) No reaction whatsoever: some authors even retain the "admin note" when replacing the submission with a new version, seemingly oblivious to its appearance. b) Attempted remediation: other authors try repeatedly to replace the submission with new versions to remove or minimize the overlapping text. (The "admin note" is retained if the amount of text overlap remains above the flagging threshold.) Some authors even politely request some form of itemization of the overlapping text, apparently unable to recall which parts of the text are original and which are reused from elsewhere. While that detailed information is not provided, some determined authors eventually succeed to eliminate the note through successive revision. c) Indignant objection: some authors have insisted that there could not possibly be text overlap (though the heuristics in place to avoid flagging false positives have proven reliable). Other authors have suggested they are following common practice, or that any overlap is inconsequential because the underlying ideas or newly intended applications are entirely different. In each of these cases, the response has been that the flagging is applied only to instances of text reuse well above the statistical background level.

**Discussion.** We first reiterate that a wide variety of full-text analyses is now technically straightforward, with established algorithms running on now-standard hardware. The arXiv database is one of the larger corpora for which the full texts are "Open Access" in the strong sense of being available for arbitrary computation. For these purposes, the larger and more comprehensive the text corpus, the richer and more accurate is the portrayal of the reuse and other behavioral patterns within research communities. Most conventional publishers understandably place restrictions on large-scale third-party harvests, so special permissions are necessary for computational analyses spanning multiple publisher databases.

Specifically regarding the text reuse analyzed here, we reiterate the lesson that the more creative and prominent authors (as measured by citation record) are typically not the offenders. We suspect that such researchers have little interest in retreading the same intellectual territory, much less reusing their own or others' material verbatim. In addition, the offending articles do not ordinarily occur in the most cutting-edge research areas, where they might be too visible, so the problem might thus be regarded as harmless to the scientific enterprise. But as we have seen, the practice is nonetheless widespread, especially in regions most vulnerable to its negative consequences. Among its pernicious effects is the fraudulent status conveyed to the perpetrators at their local institutions, and the consequent difficulty to train a next generation of researchers to break out of the cycle. The problem can be exacerbated by criteria for career advancement that reward quantity of publications without regard to their impact in the mainstream literature. The need for faux imprimatur has helped to drive the recent proliferation of "predatory open access" journals [17], which provide an additional illusion of legitimacy in the absence of expert assessment.[17] It is entirely conceivable that the problem results as much from deficiencies in educational systems and training as from willful fraudulence. In Appendix G, we consider factors which may drive text reuse by researchers, some already documented at the level of students [18].

Looking to the future, it will be informative to repeat this analysis in a few years on the arXiv corpus, to see whether the presence of the flagging has a measurable behavioral effect, or whether it simply reinforces the current behavioral norm.[18] In other words, will it have no effect on existing errant authors, will they make cynical superficial changes to evade detection, or will they make more substantive methodological changes in the way they produce research articles? While we do not expect that it will ever become acceptable to portray third party material as one's own (even if produced collectively, such as Wikipedia articles), it is possible that widespread network availability of background materials and their ease of reuse will ultimately alter the way research articles are produced, making the research enterprise more efficient by reducing redundant effort. Adaptation to the recent dramatic changes in scholarly communications infrastructure will have significant implications for how the next generation of researchers is trained, and large-scale textual analysis will continue to provide a window into how their normative behavior evolves.

**ACKNOWLEDGMENTS.** We thank Simeon Warner for improvements to the software used in [1], Scott Rogoff for writing additional analysis software during a CS Master's project, Isabel Kloumann for the question answered by Appendix E, and Gilly Leshed for pointing out [19]. This work was partially supported by NSF grant OCI 0926550.

[17] After the completion of this work, we discovered significantly higher rates of text reuse specifically in Computer Science articles published in predatory open-acess journals (articles largely received after the mid-2012 timeframe of the dataset analyzed here). We defer to any later work a more discipline-specific assessment of the issues.

[18] See [19]: "Design Claim 15: Publicly displaying many examples of inappropriate behavior on the site will lead members to believe this is common and expected."

# Supplemental Material for Patterns of Text Reuse in a Scientific Corpus

Daniel T. Citron, Paul Ginsparg
Dept of Physics, Cornell University, Ithaca, NY 14853

## A. Details of Winnowing Methodology

This work has been facilitated by the increased power of commodity hardware in recent years, in particular by the drop in cost of machines with tens of gigabytes of RAM. This allows fingerprints of the entire arXiv dataset to reside easily within memory, without swapping to disk. This has also enabled the overlap detection to be run on the 500-1000 new submissions and replacements each day in under a minute, as suggested in [1]. Since the summer of 2011, articles have been publicly flagged for text overlap with other articles.

A textual 'fingerprint' of each document in the corpus is precomputed as a set of winnowed $k$-grams drawn from the document. The $k$-grams in this context correspond to all ordered word sequences of length $k$ from a text. There are roughly $n$ of these for a text of length $n > k$ words (more precisely $n - k + 1$ of them). The value of $k$ is determined by the desired level of noise rejection. For example, the six-word phrase "this paper is organized as follows" appears in many tens of thousands of articles in the corpus and needs to be screened. The analysis in this article used $k = 7$, and hence was insensitive to phrases of less than seven words in length.

The $k$-grams are converted to hashes, and can be stored as keys of an index database, each pointing to a list of all the documents in which it occurs. For rapid lookups, this database should fit in RAM, so a winnowing methodology [2] is used to reduce its size. This winnowing is natural because the 7-grams overlap, and hence contain redundant information. To reduce the number of 7-gram hashes included in the database, a window size of $t > k$ is chosen. We consider all of the $n - t + 1$ windows in the document, each containing $t - k + 1$ $k$-grams that begin and end within the window. The algorithm retains from these only the $k$-gram with the smallest numerical value of the hash. That $k$-gram has a high probability (given by $(t - k)/(t - k + 1)$) of being the smallest as well in neighboring windows in which it appears, and any $k$-gram that is chosen in multiple windows reduces the overall number of hashed $k$-grams retained. In principle, this results in a small loss of sensitivity, since the algorithm is only guaranteed to find strings of at least $t$ successive words in common between two texts. Strings of length less than $t$ (but of course at least $k$) in common are found probabilistically, so may be missed.

In our implementation, the larger window size was chosen as $t = 12$, which means that each larger window contains six 7-grams of words that start and end in the window. This is less than both the mean and median sentence lengths in the corpus (20 and 18, respectively). Thus we see text overlaps starting at less than half the typical sentence length, and are guaranteed to see overlaps at two thirds of the typical sentence length. In practice, the overlapping articles of interest have multiple overlapping sequences of much longer than twelve words, so overlaps missed due to the abovementioned probabilistic detection ordinarily occur only in document pairs below the threshold for flagging. The winnowing fraction depends only on the quantity $t - k + 1$, equal to 6 for our values of $t = 12$ and $k = 7$, and results in a reduction in the number of stored 7-grams by a factor of about 3.6. The roughly 750,000 documents (33Gb of uncompressed text comprised of 6B words) of our database were thus characterized by roughly 1.6B hashes .

There is an additional reduction in the number of 7-grams, resulting from the elimination of so-called "common" 7-grams [1] specific to the corpus. Common 7-grams are those which appear in articles written by sufficiently many disjoint sets of authors that they do not signal copied text. They can be common phrases (e.g., "the rest of this article is organized"), boilerplate text such as copyright disclaimer, or standard text from the templates of certain conference proceedings and theses, and so on. These are easy to identify since they occur in large numbers of otherwise unrelated articles, and consequently have a distribution very different from that of actively copied content. The definition of common 7-grams used in the main text of this article is those 7-grams which occur in articles by at least four sets of disjoint authors. With the documents containing a given 7-gram considered as nodes of a co-author network, and nodes with at least one common co-author connected by edges, then in mathematical language common 7-grams are those whose co-author network has at least four disconnected components. This refinement is important because text, once copied from elsewhere, is sometimes repeatedly reused by the same authors in subsequent articles, and thus might masquerade as "common" unless each such connected co-author group is regarded as a single usage.

This definition means that we risk missing 7-grams that were independently copied at least three times, but this is a rare occurrence, and documents incorporating such text ordinarily have many other "uncommon" 7-grams copied as well. Removing the common 7-grams further reduced the number of hashes by roughly 4%. The resulting hash–document lookup table for this corpus resides in roughly 12Gb of RAM, no longer a substantial amount of memory by post-2010 standards, and permits many hundreds of lookups per second on inexpensive hardware.[1]

The 7-gram hashes for a given document provide a set of features insensitive to word sequences of less than 7 words, and can be effectively used to make pairwise comparisons between large numbers of documents. For a given number of overlapping 7-grams between two articles, the exact corresponding amount of text overlap[2] depends both on the details of the articles and on their fractional percentage of overlap. Articles with a large fractional overlap (typically with many hundreds of 7-grams in common, depending on the lengths of the articles) average fewer overlapping words per 7-gram. In the limit of 100% overlap, the number of shared words per overlapping 7-gram drops to the 3.6 winnowing ratio mentioned above, i.e., the average number of words per winnowed 7-gram. Articles with sparser overlap (a few tens of overlapping 7-grams or less), on the other hand, can be found to average more than seven words per overlapping 7-gram, both due to individual overlaps extending slightly beyond that detected by 7-gram hash, and due to other overlapping sequences missed by the probabilistic winnowing procedure.

[1] Parts of the methodology, as described in sec. 2 of the text, were specific to the arXiv corpus, so we have not benchmarked our methods against those reported in [3].

**Summary:** The number of hashes retained for each article is reduced by about a factor of 3.6 by the winnowing procedure [2], significantly reducing the RAM footprint of the hash database at only a small cost in sensitivity. Our choice of parameters provides a guarantee of finding any matching word sequence of at least 12 words in a row between two articles, and some (decreasing) probability of detecting sequences of length less than 12 (and at least 7) words in a a row in common. The hashes retained are further reduced by eliminating "common" 7-grams [1], here defined as those appearing in articles written by at least four disjoint sets of authors, and thus not likely to indicate copied text.

[2]calculated using the actual text, rather than the winnowed 7-grams

## B. Some Guidelines

While there is no universal standard pertaining to reuse of text in scientific publications, many universities and publishers have established explicit guidelines regarding publishing articles with reused text. In addition, many universities, including our own [4], provide guidelines or some form of training for avoiding unethical behavior, including plagiarism. They typically point to materials at the Federal Office of Research Integrity [5], which state that "Substantial unattributed textual copying of another's work means the unattributed verbatim or nearly verbatim copying of sentences and paragraphs which materially mislead the ordinary reader regarding the contributions of the author." These also clarify that use of common phrases within a community is not considered misleading, and a finding of misconduct generally requires a "significant departure from accepted practices of the relevant research community." The Research Councils UK [6] specifies both "misappropriation or use of others' ideas" and "undisclosed duplication" of one's own work as unacceptable publishing practice. We have not conducted a systematic check of policies maintained by federal offices and research oversight agencies in other countries, but note that journal policies do provide effective international guidelines.

The American Physical Society's guidelines [7] for submissions to its journals are unequivocal regarding text reuse: "Authors may not ... incorporate without attribution text from another work (by themselves or others), even when summarizing past results or background material. If a direct quotation is appropriate, the quotation should be clearly indicated as such and the original source should be properly cited." These guidelines do permit "material previously published in an abbreviated form" to provide the basis for a more detailed article, as long as reproduction of previously used material is minimized and properly referenced. We have seen that materials submitted to the arXiv do not always conform to these exacting standards, and yet are published by journals, indicating that editors do not systematically employ an automated screen.

## C. Other Sample Authors

To give a flavor for some of the behavior of individual authors in the dataset, fig. S1 shows the text overlap graphs for two authors exhibiting extreme frequency of text reuse. The visualized text overlap networks use the same convention as described for fig. 3. These are relatively prolific authors whose articles reuse large blocks of text both from their own previous articles and from articles they cite. The paucity of red-colored links indicates that at least neither of these authors appears to draw heavily from uncited articles by others.

These two examples are also representative of another phenomenon we have noticed: the existence of groups of authors who have self-organized into sub-communities based on their frequent copying of text from, and citations of, one another. The present study focuses first on global measures of text reuse and detection of single individuals with anomalous behavior.

a)

b)

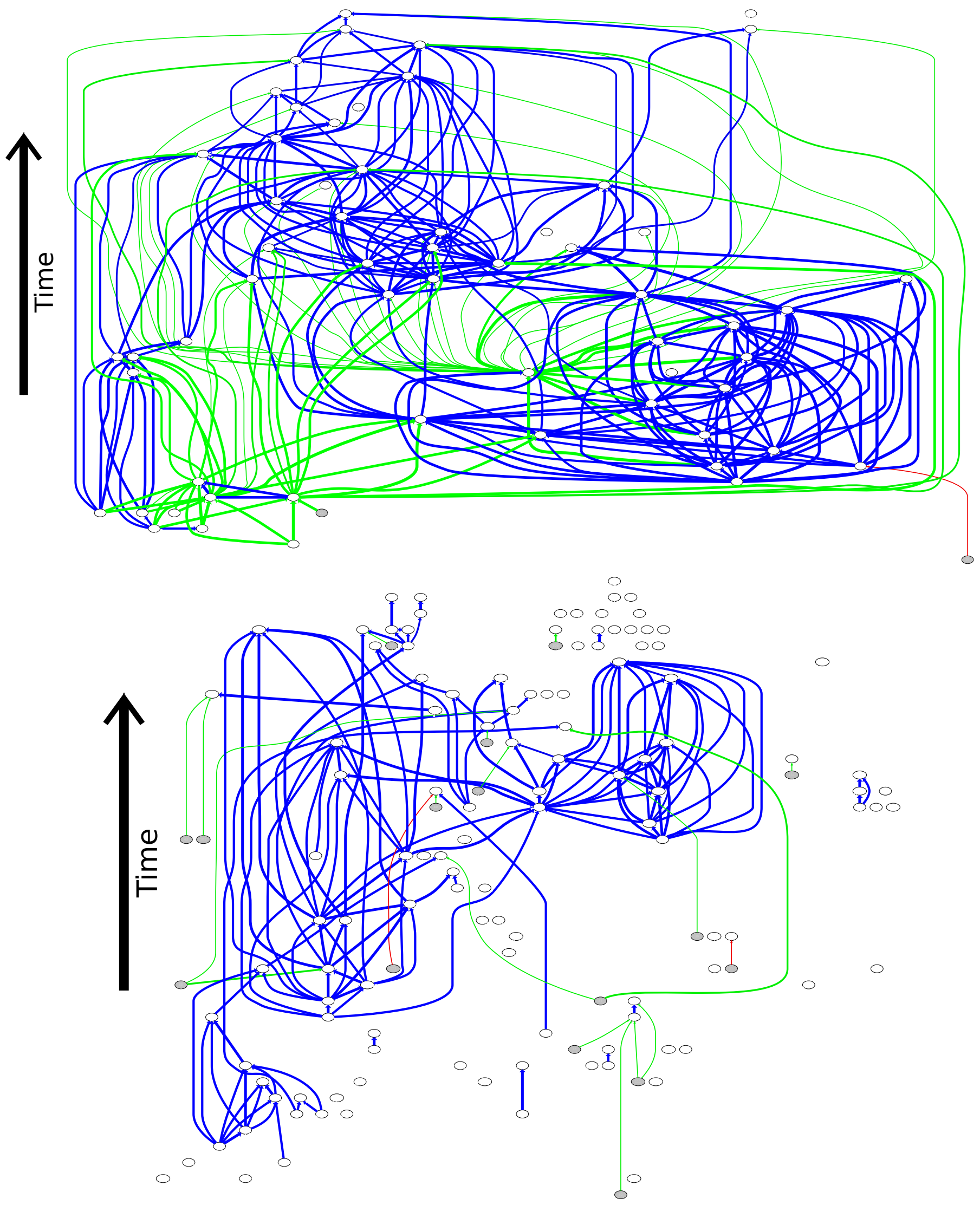

Figure S1: Overlap graphs for two authors. The number of articles and timeline of publication for each is a) 65 from 2003 to 2012, b) 126 from 1999 to 2012. As elsewhere, the blue lines indicate significant self-copying (AU), the green lines reuse from other authors with citation (CI), and the red lines reuse without citation (UN), above the thresholds of 100, 20, and 20 common 7-grams, resp.

We also find many cases of groups of authors who not only frequently copy from themselves but also from one another. These authors' articles group together in the visualizations of their text overlap networks, separated from the bulk of other authors. Some co-authors of the two authors depicted in fig. S1 have quantitatively similar overlap networks. The behavior that leads to the formation and growth of these author groups warrants further study.

# D. Some special cases

In this appendix, we give some examples of the subtleties involved in classifying instances of text reuse. Each represents known cases for which the overlap detection scheme has found

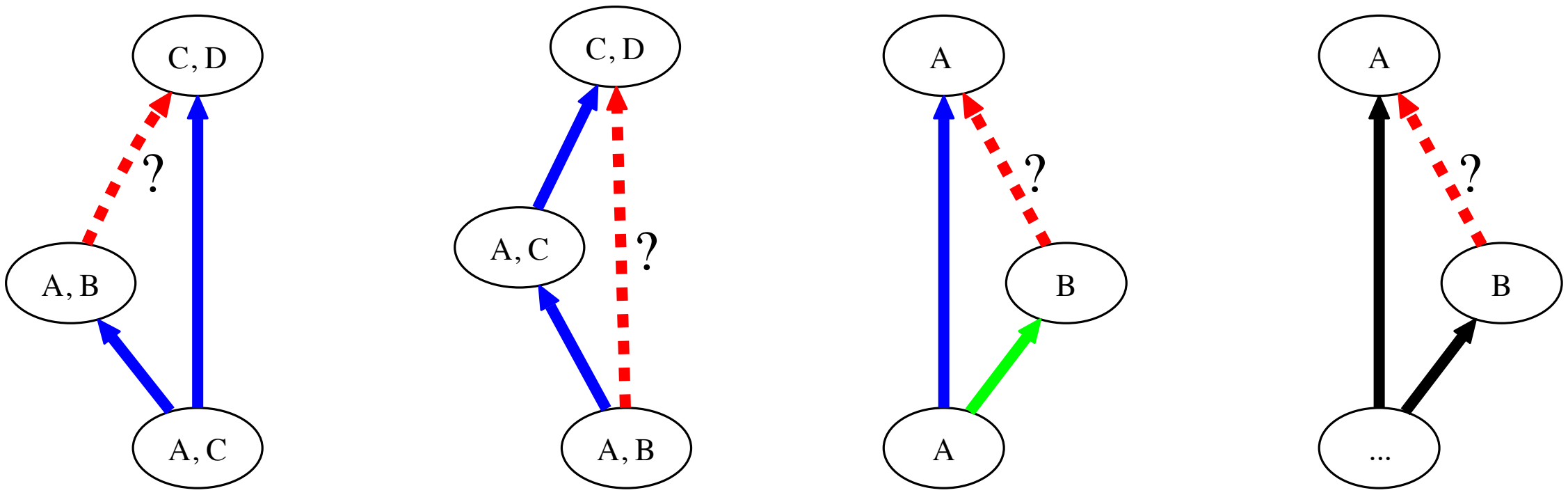

a) b) c) d)

Figure S2: a) overlap between articles by disjoint authors may be due to a prior coauthored article. b) reuse of text by author C that was demonstrably not written by author C, but whose propriety falls within a gray area. c) A seems to be reusing text from B, but actually is reusing own text. d) A seems to be reusing text from B, but actually both are reusing from a common source.

false positives.

Fig. S2a shows a common case in which an article by authors C,D might be flagged as having inappropriate overlap with an article by authors A,B (red dotted line). But authors A,C had previously written an article together, so that in principle authors A and C each feel entitled to reuse a block of text from a previous article (though of course only one of them might have written the original text in question).

Fig. S2b shows a similar case, in which an article is written by authors A,B and a block of text is subsequently reused in an article by A, now coauthoring with C. Author C then goes on to reuse the same block of text in an article coauthored with D. (B and C might be students working with A.) Once again, there appears to be inappropriate overlap (red dotted line) between the article by authors C,D and the one by A,B. Author C might claim that it is simply being reused from the earlier co-authored article with A, except in this case it can be demonstrated that the text was not written by C (or D), since it had appeared in an earlier article by A,B without C. There are many cases of this type in the arXiv dataset (though not as many as case a)), and it is not clear whether they should be classified in some gray area between self-copying and copying from cited sources.

Fig. S2c shows a case in which author A might appear to be reusing text from author B without citation (dotted red line), but it is actually author B at fault, having reused text from an earlier article by A. This case can be problematic to detect if the earlier article by A is not in the arXiv dataset.

Fig. S2d shows a generalization of case c) in which authors A,B independently use text from some earlier article, perhaps by other authors entirely, and perhaps not in the arXiv dataset. While author A might appear to be inappropriately reusing text from B without citation, it might be that both are reusing text from an earlier source, either appropriately (in quotes and cited) or inappropriately.

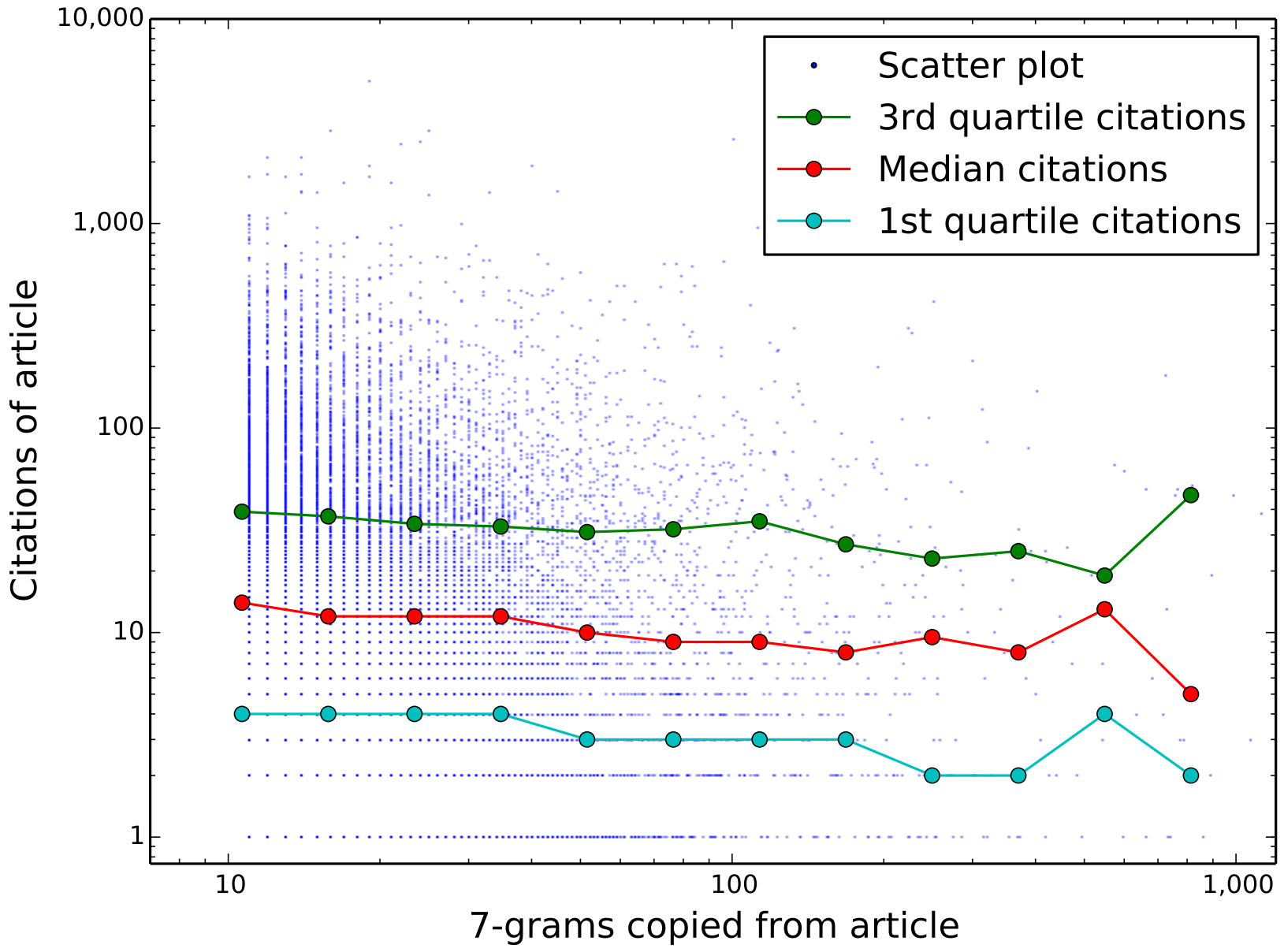

Figure S3: The median, first, and third quartiles for the citation data on a logarithmic scale. Each scatter plot data point (blue) represents the number of citations that an article has received, given a single instance of Cited or Uncited text reuse from that article. The median number of citations vs. amount of text reused is shown in red, indicating a negative correlation, with Spearman correlation coefficient $r = -.625$ ($p = .030$). The anomalous data point at the middle right may may be exaggeratedly high due to relatively few instances of very large amounts of copying.

# E. What gets copied?

Here we use the citation data to assess the relative importance of articles that are used as sources for borrowed or reused text. Again, we use citation count as a proxy measure for influence of an article. The dataset used is the same as in sec. 5.1: we consider cases of Cited and Uncited reuse of text from those articles for which we have more complete citation data, and exclude self-citations. (Cases of Common Author text reuse have already been shown to receive fewer citations and would bias the results.)

Fig. S3 shows the correlation between citations of articles and instances of text reused from those articles. Each data point in the scatter plot data represents a single instance of text reused from an article. (For example, there is a single data point for an author who reuses 100 7-grams from an article that has received 5 citations. Multiple cases of text reuse from a single article result in multiple data points.) The figure shows the decreasing trend of median number of citations as the amount of reused text increases, and the negative correlation has a Spearman coefficient $r = -.625$ ($p = .030$). We see that authors tend to reuse content from articles that receive less mainstream recognition from the research community.

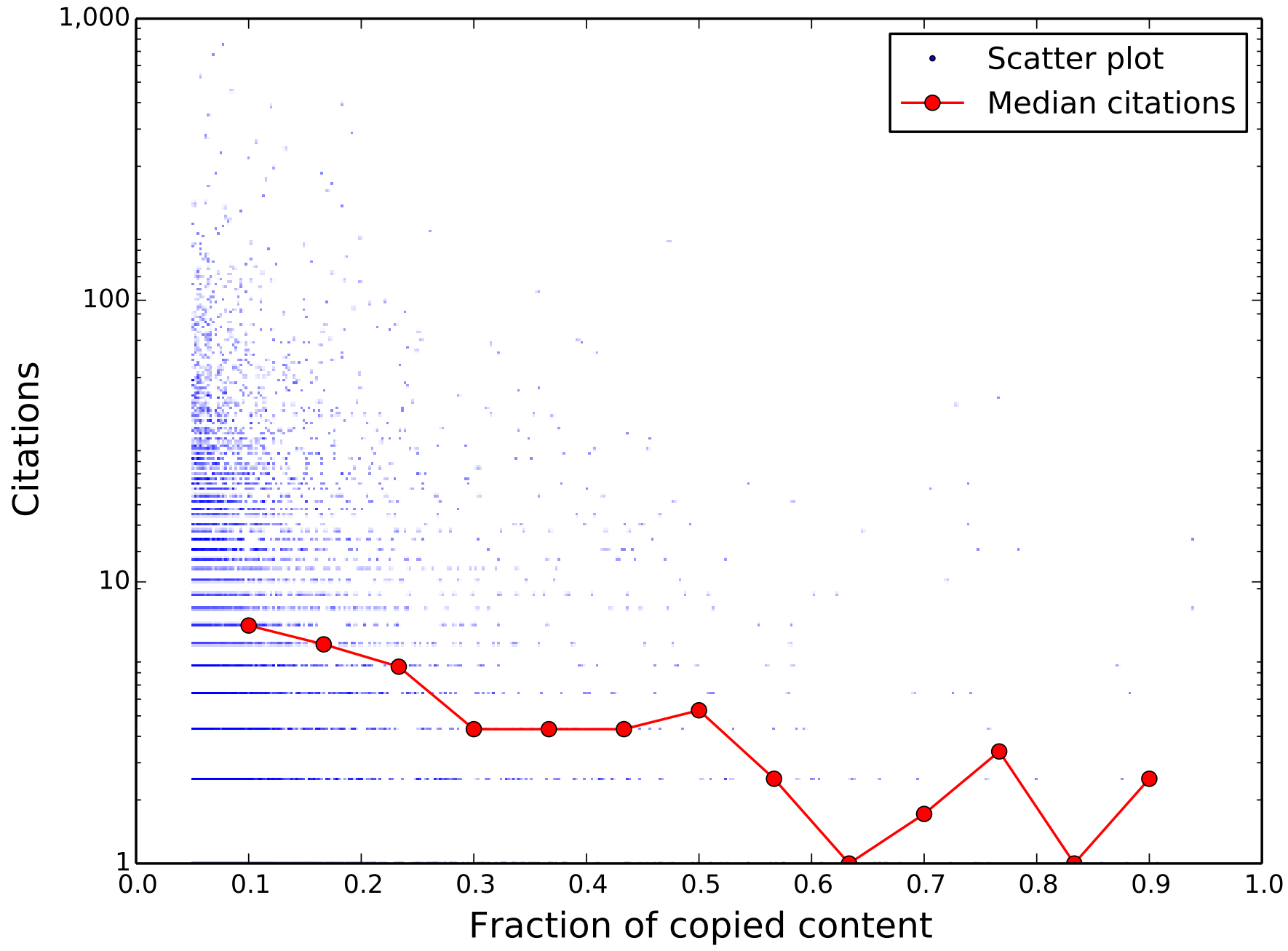

Figure S4: Data restricted to articles submitted from Germany, France, and Italy, three countries with relatively low rates of copied content. Each scatter plot data point (blue) represents the number of citations that an article has received, given a single instance of Cited or Uncited text reuse from that article (see fig. 5). The median number of citations vs. amount of text reused is shown in red, indicating a negative correlation between the number of citations and the amount of copied content, with Spearman correlation coefficient $r = -.859$ ($p = 1.71 \cdot 10^{-4}$).

# F. Country of origin bias

We argue here that the negative correlation between the number of citations received by an article and its amount of copied text (seen above in fig. 5, and discussed in sec. 5.1) is not biased by the article's country of origin. The concern is that, as seen in sec. 5.2, some countries produce articles with significant amounts of copied text at higher rates than others. If articles from those countries already accrue fewer citations on average, then the negative correlation in fig. 5 may be caused by that prior correlation.

To exclude this effect, we restrict attention in fig. S4 to articles from three countries with relatively low rates of copied content: Germany, Italy and France (6307 articles total). We see that the effect persists: even for countries with low rates of text reuse, there is still a negative correlation between the number of citations and the amount of text copied. The negative correlation is also present in the data individually for each of these countries.

## G. Some Underlying Causes?

As mentioned in sec. 6 of the text, it is entirely conceivable that the problem of excessive text reuse by some authors results as much from deficiencies in educational systems and training as from willful fraudulence. One indication of this is the relative rarity of uncited reuse of text: that many researchers readily include blocks of text (though not within quotation marks) from cited sources suggests that they really do regard this as common practice, and have nothing to hide. Similarly, they may believe that series of articles by the same authors are ordinarily produced from the same template, with large successive overlaps, as a standard practice. Producing a new idea is an act of magic, requiring substantial experience and training, and without correct mentoring it may be taken for granted that articles are instead produced by weaving together texts from existing sources. While conceivably exacerbated by the ease of cutting and pasting text in electronic format, the problem does predate both the new technology and the use of preprints. Ironically the combination of those make make that reuse that much easier to detect.

Studies of international students have suggested significant cultural differences in attitudes toward plagiarism [8], stemming from different views of the written word [9]. Many students from non-western cultures had never before heard the word plagiarism, and in some cultures it is considered disrespectful to rewrite another author's words. The undergraduate students interviewed in [8], who thought it acceptable to copy material from the internet, or to plagiarize what they regarded as very general or background information, would need specific additional training before joining the international research community. Yet an additional mitigating factor is that non-native speakers of English may see little point to translating "standard" or "background" material internally into their native language, and then back into English, and indeed the problem documented in arXiv submissions is more prevalent among such non-native speakers writing in English (see also [10]). This too was evident at the student level [8, 9], where in order to explain concepts, students less confident in their English proficiency tended to employ longer phrases from other sources, rather than just words. At a later career stage, a desire to publish in higher quality venues would provide additional impetus in this direction. A researcher concerned that his or her articles are rejected due to the quality of writing may feel compelled to imitate sentence structures from other articles.